\documentstyle[aps]{revtex}

\begin{document}

\title{Possible critical point in phase diagrams of interlayer
Josephson-vortex systems in high-\(T_c\) superconductors}

\author{Xiao Hu and Masashi Tachiki}

\address{National Research Institute for Metals, 
Tsukuba 305-0047, Japan}

\date{submitted to PRL on Feb. 11, 2000}

\maketitle

\begin{abstract} 
A critical value in the product of the anisotropy parameter and the 
magnetic field is observed in interlayer Josephson-vortex systems by 
extensive Monte Carlo simulations.  Below/above this critical value the
thermodynamic phase transition between the normal and the
superconducting
states upon temperature sweeping is first/second order. 
The origin is the intrinsic pinning effect of the 
layered structure of high-\(T_c\) superconductors. 

\vskip1cm

\end{abstract}

\newpage 

In the superconducting state, an external magnetic field applied 
in the Cu-O plane of a high-\(T_c\) superconductor induces the
so-called Josephson vortices.  The center of a Josephson vortex
enters into a block layer, the layer between two superconducting
Cu-O layers, in order to save the condensation energy of 
superconductivity \cite{Tachiki}.  
The thermodynamic phase transition and the lattice structure of
interlayer Josephson vortices have been attracting considerable 
interests since the discovery of high-\(T_c\) superconductivity.
Using a London theory, the structure of Josephson-vortex lattice 
was derived as the compressed hexagons of triangular lattice pointing 
along the c axis by Ivlev, Kopnin and Pokrovsky \cite{Ivlev}. 
The interlayer shear modulus was shown to be exponentially small, 
and the shear deformation of a rhombic lattice might arise through a second-order phase transition.  However, at higher temperatures 
fluctuations are more important, and the London theory is generally 
inaccurate for discussions about phase transitions. Considerable efforts 
have been made in order to clarify the thermodynamic
phase transition in Josephson-vortex systems both experimentally
\cite{Kwok,Schilling,Schmidt,Ishida,Hussey} and theoretically 
\cite{Blatter,Mikheev,Horovitz,Balents,Hu2} thereafter.
Nevertheless, the understanding for the problem is still not satisfactory
yet.  Results obtained by different techniques even seem to be
contrary to each other.  The difficulty in approaching this problem is 
two-fold.  On the experimental side, a small deviation of the direction 
of the magnetic field from the Cu-O plane can lead to a strong 
influence of 
the c-axis component of the magnetic field on thermodynamic properties 
of the systems, since all the high-\(T_c\) superconductors are very 
anisotropic.  On the theoretical side, one has to treat simultaneously 
anisotropic inter-vortex forces, the commensuration between
the vortex alignment and the underlying layered structure, and 
thermal fluctuations.

In the present Letter, we report new results of extensive
Monte Carlo (MC) simulations on the thermodynamic phase transition and
the lattice structure of interlayer Josephson vortices.  Our
results suggest the existence of a critical value of product of the
anisotropy parameter and the magnetic field, such that below/above this 
critical value the thermodynamic phase transition between the normal 
and the superconducting states is first/second order upon temperature
sweeping.

The model Hamiltonian used for the present simulations is the 
so-called 3D anisotropic, frustrated XY model defined on the
simple cubic lattice \cite{Huse,Teitel,Hu1,Sudbo1}:  

\begin{equation}
  H=-J\sum_{<i,j>\parallel x {\rm axis}} 
                     \cos(\varphi_i -\varphi_j)
            -J\sum_{<i,j>\parallel y {\rm axis}} 
                     \cos(\varphi_i-\varphi_j) 
        -\frac{J}{\gamma^2}\sum_{<i,j>\parallel c {\rm axis}} 
                     \cos(\varphi_i-\varphi_j
                              -\frac{2\pi}{\phi_0}\int^{j}_{i}A_cdr_c).
\end{equation}

\noindent Here the y axis is along the external magnetic field, and 
\({\bf y}\perp {\bf c}\perp {\bf x}\). The unit length of the simple cubic 
lattice is the distance \(d\) between the neighboring Cu-O layers in a cuprate.  Therefore, the discreteness in the c axis comes from
the underlying layered structure of cuprates, while that in the
Cu-O planes is introduced merely for computer simulations. 
The coupling constant is given by 
\(J=\phi^2_0 d/16\pi^3\lambda^2_{ab}\). The anisotropy parameter is 
defined by \(\gamma=\lambda_c/\lambda_{ab}\), and determines the 
ratio between the couplings in the Cu-O plane and along the c axis.  In the 
present model, fluctuations in amplitudes of superconducting order 
parameters and in the magnetic induction are neglected. 

Details of simulation technique are summarized as follows: The density 
of flux lines induced by the external magnetic field is 
\(f=Bd^2/\phi_0\).  A Landau gauge is adopted so that 
\(A_x=0\) and \(A_c=-xB\). The system size is 
\(L_x\times L_y\times L_c=384d\times 200d\times 20d\), which is 
compatible with the filling factor \(f=1/32\).  There are 240 Josephson flux lines in the ground state. Periodic boundary 
conditions are applied on phase variables in all directions.
A typical simulation process is started from a random configuration of 
the phase variables at a high temperature, such as \(T=1.5J/k_B\). 
30000 and 90000 MC sweeps are used for equilibration and statistics,
respectively, at each temperature.  The last configuration at a 
temperature is used as the initial configuration at a slightly lower 
temperature, where the temperature difference is 
\(\Delta T=0.1J/k_B\). Around the transition temperature, more than one 
million MC sweeps are adopted at each temperature, and 
meanwhile the cooling rate is reduced to \(\Delta T=0.01J/k_B\).  
Vortices are identified by counting phase differences around plaquettes. 

In order to compare our simulation results with existing experimental 
observations, we choose to study first a system of anisotropy parameter 
\(\gamma=8\), which is near to  that of 
YBa\(_2\)Cu\(_3\)O\(_{7-\delta}\).  The magnetic field corresponding to
\(f=1/32\) in our simulations is much stronger than those in 
experiments, and we come back to this point later.
The temperature dependence of the helicity modulus (a quantity 
proportional to the superfluid density) along the magnetic 
field and the specific heat is depicted in Fig. 1. There is a 
clearly observable \(\delta\)-function like peak in the specific heat 
at \(T_m\simeq 0.96 J/k_B\), where the helicity 
modulus along the direction of magnetic field
increase sharply from zero.  Shown in the same figure
is the temperature dependence of the intensities of Bragg peaks in diffraction patterns at 
\({\bf q}^{(1)}_{xc}=(\pm \pi/8d, 0)\) and 
\({\bf q}^{(2)}_{xc}=(\pm \pi/16d, \pm \pi/d)\). 
Therefore, a thermodynamic first-order phase transition occurs
at \(T_m\), where the gauge symmetry and translation symmetry 
are broken simultaneously, corresponding to the realization of
superconductivity and Josephson-vortex lattice respectively.

The lattice structure of Josephson vortices at low 
temperatures is shown in Fig. 2.  The unit cell is rhombic with
short axis along the c direction and of a length of \(2d\), and the long 
axis along the x direction and of a length of \(32d\).  Josephson vortices 
are distributed in every block layer for the present parameters
\(\gamma=8\) and \(f=1/32\). This structure is the same as that 
predicted by Ivlev, Kopnin and Pokrovsky \cite{Ivlev}.  

The lattice structure in Fig. 2 is obviously the ground state for 
\(\gamma\ge 8\) when the filling factor is fixed at \(f=1/32\). 
Therefore we can use it for investigations of thermodynamic 
properties for large anisotropy parameters by a heating process.  
The specific heats thus obtained are shown in Fig. 3 for anisotropy
parameters \(\gamma=8\), 9, and 10.
The \(\delta\)-function peaks in the curves for \(\gamma=8\) and 9,
is suppressed for \(\gamma=10\) \cite{Hu2}.  
In Fig. 4 we display the
temperature and anisotropy parameter dependence of the phase
difference between nearest neighboring Cu-O layers
\(\langle \cos(\varphi_n-\varphi_{n+1})\rangle\).
There is a jump in \(\langle \cos(\varphi_n-\varphi_{n+1})\rangle\) 
for \(\gamma=8\) and 9, which is smeared out for
\(\gamma=10\).  As the jump in 
\(\langle \cos(\varphi_n-\varphi_{n+1})\rangle\) is nothing but the 
jump in the Josephson energy in units of \(J/\gamma^2\),  there exists a 
latent heat at the transition temperature for \(\gamma=8\) 
and 9, but not for \(\gamma=10\), consistently with the 
data for the specific heat.  
The value of the latent heat itself is too tiny, about \(\gamma^2\) 
times smaller than that in 
\(\langle \cos(\varphi_n-\varphi_{n+1})\rangle\), 
to be detected directly.  On the other hand, from a standard finite-size
scaling theory for a first-order phase transition, the height of the
\(\delta\)-function like peak in the specific heat is proportional to 
the system size \cite{Hu1}.  Therefore, by using a large system such 
as the one in our simulations, the \(\delta\)-function like peak in the 
specific heat becomes observable as in Figs. 2 and 3 for the first-order
phase transitions. 

We have performed simulations for anisotropy parameters
\(\gamma=7, 6,\cdot\cdot\cdot\), down to the isotropic case of
\(\gamma=1\) \cite{Brezin}
fixing the filling factor at \(f=1/32\), and observed 
first-order phase transitions for all these anisotropy parameters.  
Therefore, the present simulation
results indicate that there is a critical anisotropy parameter in
between \(\gamma=9\) and \(\gamma=10\) for \(f=1/32\), 
below/above which the phase transition is first/second order.

Now we look for the reason of the suppression of the
first-order phase transition when the anisotropy parameter is increased.
Suppose a complete commensuration is achieved between the alignment 
of the Josephson vortices shown in Fig. 2 and the underlying layered 
structure of high-\(T_c\) cuprates.  In other words, the Cu-O layers
do not influence the lattice structure of Josephson vortices,
but merely fix its position in the c direction.  In such a case, the 
Josephson-vortex
lattice should be rescaled into equilateral triangular lattice using the
anisotropy parameter \(\gamma\), and we have a relation as seen
in Fig. 5:

\[
 (2d)^2=d^2+(d/2f\gamma)^2, \nonumber
\]

\noindent which results in

\begin{equation}
 f\gamma =\frac{1}{2\sqrt{3}}.
\end{equation}

\noindent Now we increase the anisotropy parameter from that
determined by the above relation when the filling factor \(f\) is fixed.
Since the repulsive force between Josephson vortices in the c direction 
is reduced, the Josephson-vortex lattice would be compressed in this
direction in order to achieve the energy minimum for the new anisotropy
parameter.  This reconstruction of Josephson-vortex lattice is
forbidden by the underlying layered structure of the high-\(T_c\) 
superconductor.  Therefore, the above relation provides a criterion for 
onset of 
the intrinsic pinning effect of the layered structure on the formation 
of Josephson-vortex lattice.  For anisotropy parameters larger than
that evaluated by the above relation, the lattice structure
 of Josephson vortices
is determined by both the inter-vortex repulsions and the pinning force 
of the underlying layered structure.  The thermodynamic phase transition
associated with the formation of the Josephson-vortex lattice can be 
different in the two regions divided by the above relation.  

Numerically, the critical anisotropy parameter for the filling factor
\(f=1/32\) is evaluated as \(\gamma=16/\sqrt{3}\simeq 9.24\) by 
the relation (2).  This estimate
coincides well with our simulation results, since  
first-order phase transitions are observed for 
\(\gamma\le 9\) but not for \(\gamma\ge 10\).  We have also performed
simulations for the filling factor \(f=1/25\), and found the variation
of phase transition from first to second order around \(\gamma=8\).  
This observation is consistent with the relation (2), since
for \(f=1/25\) one has the critical anisotropy parameter
\(\gamma\simeq 7.22\). For \(f=1/36\), we have observed a first-order
phase transition even for \(\gamma=10\), consistently with the 
critical value \(\gamma\simeq 10.39\).  Namely, our simulation results
indicate clearly that the critical anisotropy parameter increases with 
decreasing filling factor, or magnetic field.  Quantitatively, the
simple relation (2) seems to give a reasonable estimate on the critical 
anisotropy parameter.

The same variation of the phase transition should be observed when the 
anisotropy parameter is fixed while the filling factor, or the strength of the magnetic field, is tuned.  The relation (2) can be rewritten as

\begin{equation}
 B=\frac{\phi_0}{2\sqrt{3}\gamma d^2}.
\end{equation}

\noindent For YBa\(_2\)Cu\(_3\)O\(_{7-\delta}\) 
with  \(\gamma\simeq 8\) and \(d=12\AA\), the critical magnetic field
is estimated as \(B\simeq 50T\). Therefore the phase transition in the 
Josephson-vortex systems in YBa\(_2\)Cu\(_3\)O\(_{7-\delta}\)  is 
first order for magnetic fields available experimentally, according to
our present study. 
For Bi\(_2\)Sr\(_2\)CaCu\(_2\)O\(_{8+y}\) with  
\(\gamma\simeq 150\) 
and \(d=15\AA\), the critical magnetic field is evaluated as 
\(B\simeq 1.7T\),
which can be checked experimentally.

The phase transition of interlayer Josephson vortices in a layered 
superconductor has been addressed theoretically by Blatter, Ivlev and 
Rhyner \cite{Blatter} and Balents and Nelson \cite{Balents}.  
In these theories, the Josephson-vortex lattice melts in-between
the layers with the formation of a smectic-like vortex liquid,
via a second-order phase transition (see also \cite{Ivlev}).  
Therefore, for strong magnetic fields the theories give a reasonable 
scenario for the phenomena
in Josephson-vortex systems observed in the present simulations.

The importance of thermal fluctuations in the phase transition should be 
stressed.  For example, about one vortex-antivortex pair is thermally 
excited per Josephson flux at each xc section at \(T=0.8J/k_B\),
a temperature lower than the corresponding transition point, for
\(\gamma=8\) and \(f=1/32\)  \cite{Hu4}.  Energetically, an additional 
Josephson vortex only costs energy of order of \(J/\gamma^2\), which 
becomes very small when the anisotropy parameter \(\gamma\) is large
\cite{Sudbo2}. 
Most of the thermally excited Josephson vortices and antivortices are 
confined in same block layers and form overhangs in flux lines, or closed loops.  As the result, for large anisotropy parameters Josephson flux 
lines induced by the magnetic field collide with each other in same
block layers even below the transition temperature \cite{Hu4}.  Therefore, these thermally excited Josephson vortices and antivortices 
play important roles in smearing out the first-order phase transition in
a Josephson-vortex system in a layered superconductor with a large 
anisotropy parameter \cite{Mikheev,Nono}.

In summary, from extensive Monte Carlo simulations we have
found a critical value in the product of the anisotropy parameter and 
magnetic field in interlayer Josephson-vortex systems in 
high-\(T_c\) superconductors.  Below/above this critical value,
the thermodynamic phase transition between the normal state
and the superconducting state is first/second order.  
According to the present results,
the phase transition in Josephson-vortex systems in 
YBa\(_2\)Cu\(_3\)O\(_{7-\delta}\) is first order under magnetic fields 
up to \(B\simeq 50T\), while it varies from first to second order in 
Bi\(_2\)Sr\(_2\)CaCu\(_2\)O\(_{8+y}\) as the magnetic field is 
increased to across \(B\simeq 1.7T\).

The authors thank L. Bulaevskii, J. Clem, A. Sudb\o \hskip3mm and 
Y. Nonomura for useful discussions.  
The present simulations are performed on the Numerical Materials 
Simulator (SX-4) of National Research Institute for Metals (NRIM), Japan.


\vskip5mm


\noindent Fig. 1:  Temperature dependence of the helicity modulus
along the magnetic field, the specific heat and the intensities 
\(I_1\) and \(I_2\) for the Bragg peaks at 
\({\bf q}^{(1)}_{xc}=(\pm \pi/8d,0)\) and 
\({\bf q}^{(2)}_{xc}=(\pm \pi/16d,\pm \pi/d)\) respectively for 
\(\gamma=8\) and \(f=1/32\).

\noindent Fig. 2:  Josephson vortex lattice for \(\gamma=8\) and 
\(f=1/32\) obtained by MC simulations of a cooling process from a 
random state at high temperatures.

\noindent Fig. 3: Temperature dependence of the specific heat for 
\(\gamma=8\), 9 and 10 and \(f=1/32\).  Data for \(\gamma=8\) and 9 
are shifted by constants.

\noindent Fig. 4: Temperature dependence of the phase difference
between nearest neighboring Cu-O layers  for \(f=1/32\).  The lines are 
for eye-guide.

\noindent Fig. 5: Real-space unit cell of the Josephson-vortex lattice 
in a layered superconductor of a large anisotropy parameter.

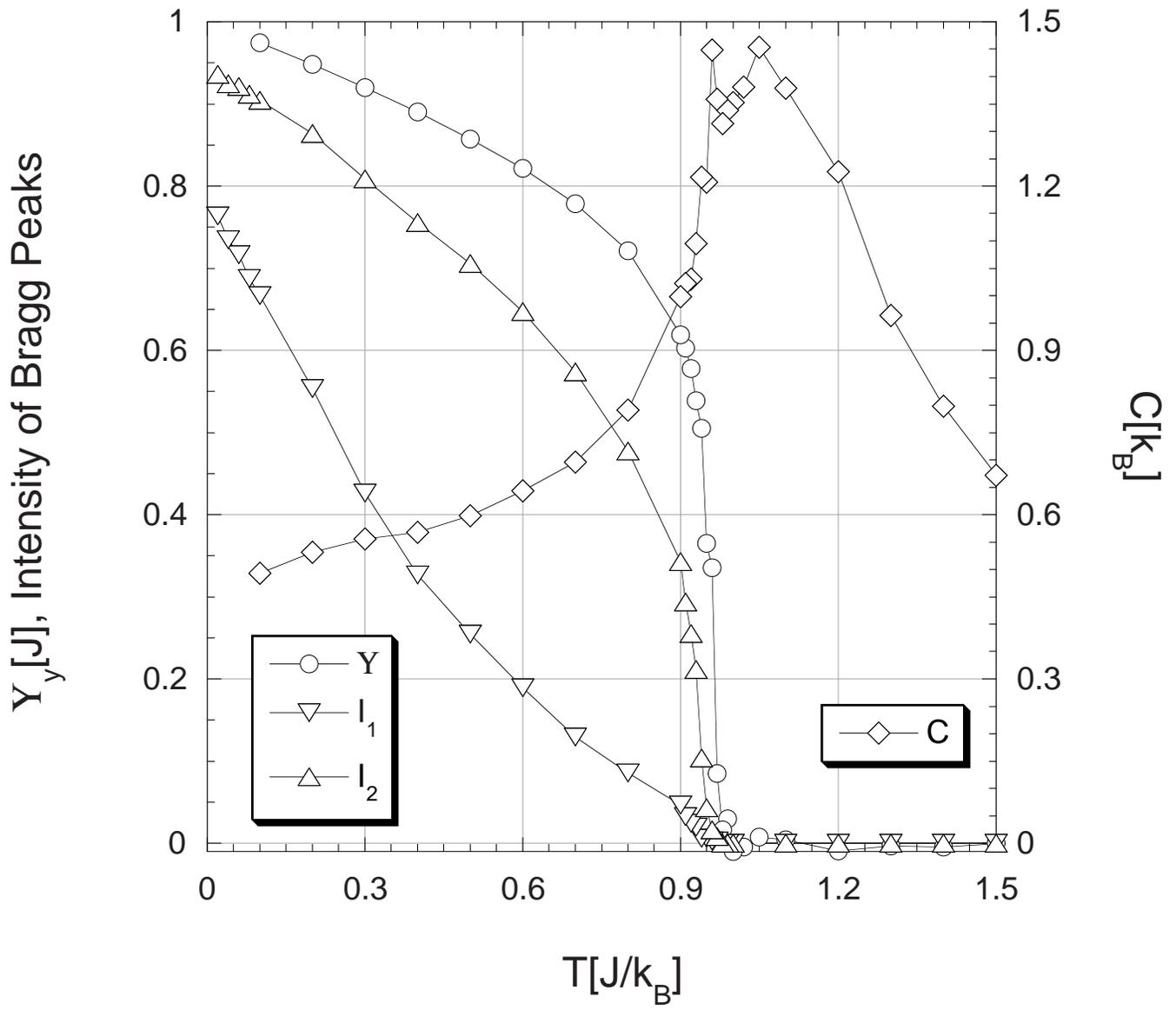
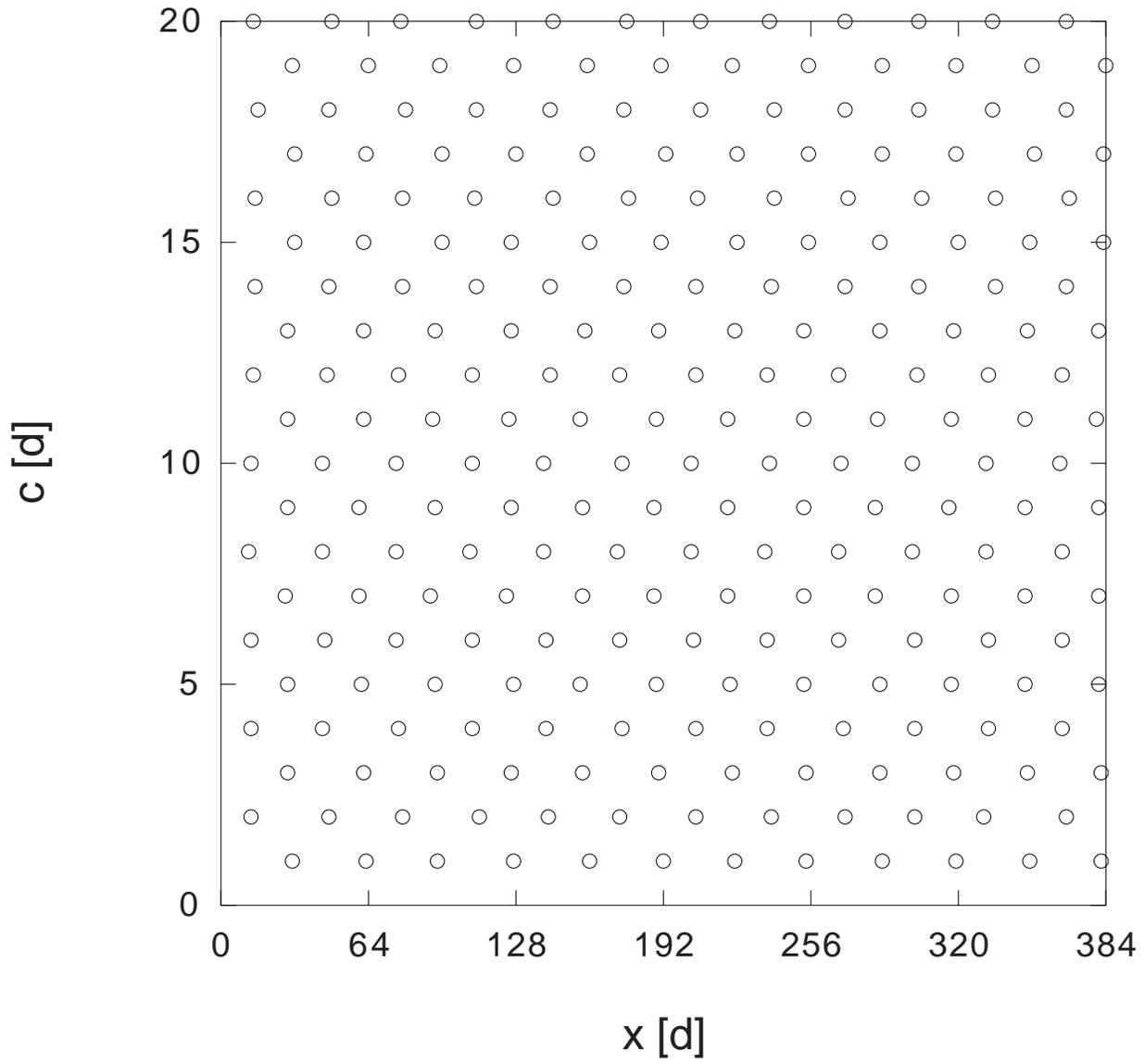
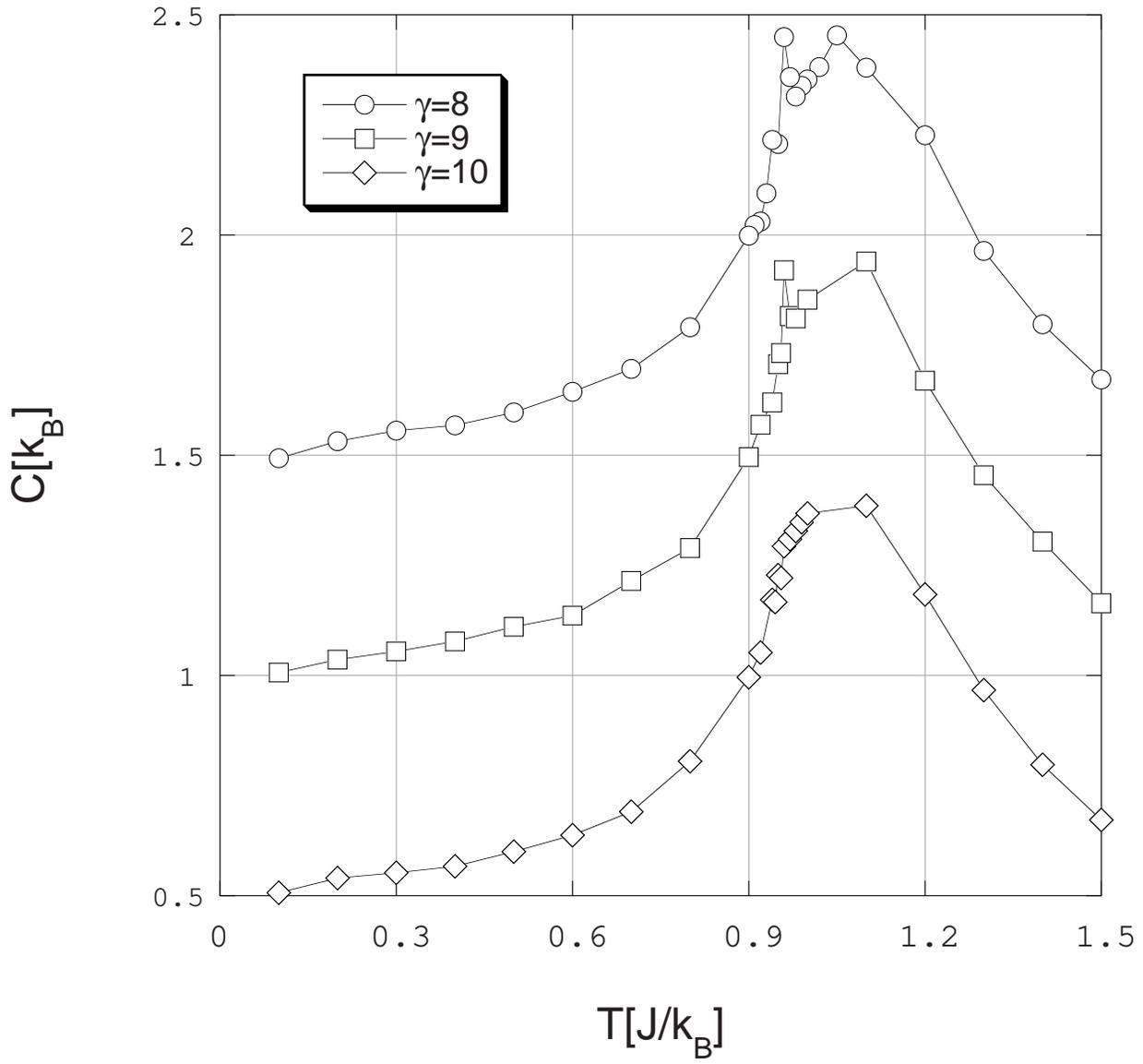
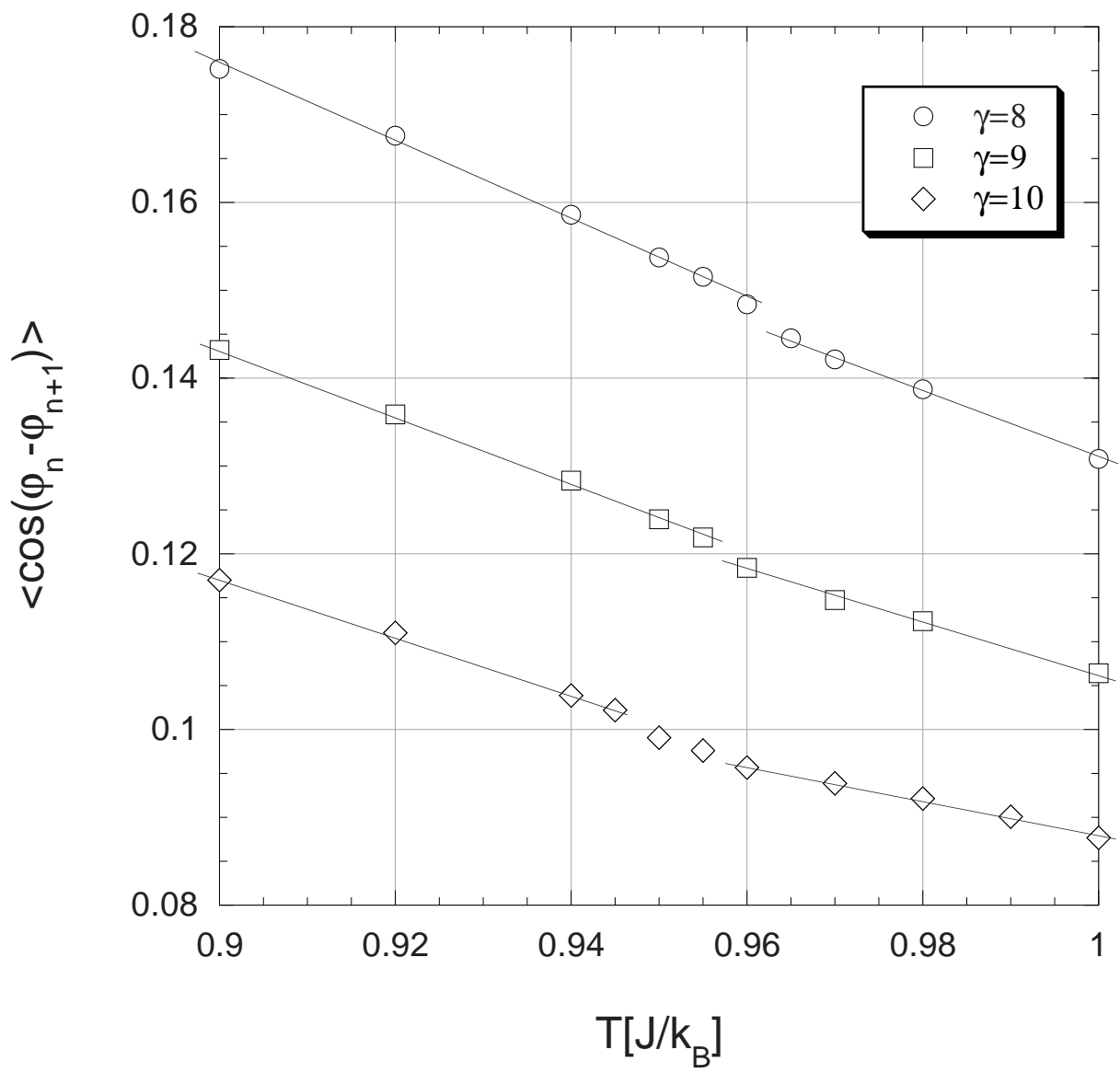
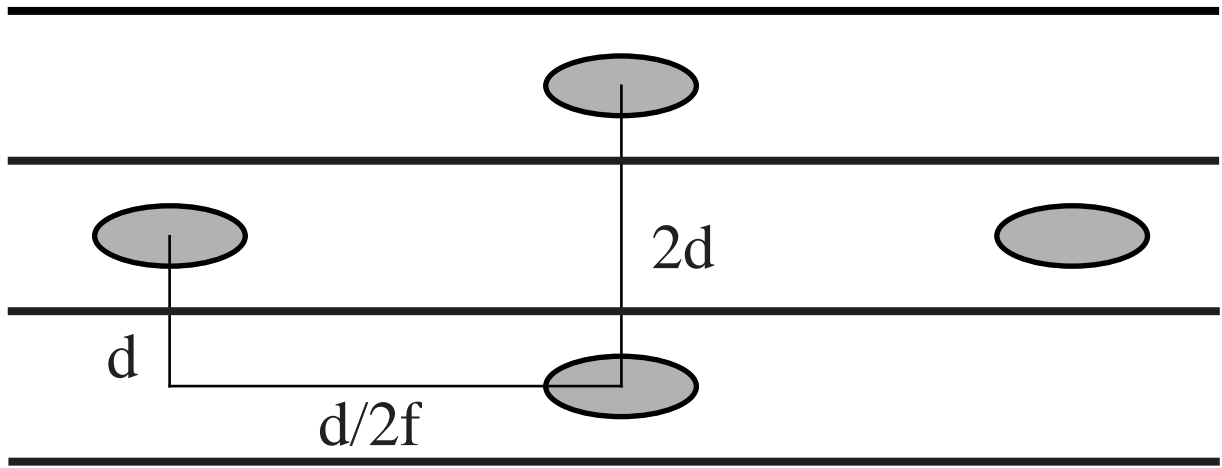
\end{document}